\documentclass[12pt]{article}
\usepackage{geometry}                
\def\calb{{\cal B}}
\def\calc{{\cal C}}

\def\calr{{\cal R}}
\def\calw{{\cal W}}
\def\calrk{{\cal R}^{(k)}}

\def\calg{{\cal G}}
\def\cals{{\cal S}}

\def\b2hat{ {\hat b}_2 }

\begin{document}

\begin{titlepage}
\vfill
\begin{flushright}
\end{flushright}

\vfill
\begin{center}
\baselineskip=16pt
{\Large\bf   Conformal Tensors via Lovelock Gravity}

\vskip 0.15in
\vskip 0.5cm
{\large {\sl }}
\vskip 10.mm
{\bf 
David Kastor \\

\vskip 1cm
{
	Department of Physics, University of Massachusetts, Amherst, MA 01003\\	
	\texttt{kastor@physics.umass.edu}
     }}
\vspace{6pt}
\end{center}
\vskip 0.2in
\par
\begin{center}
{\bf Abstract}
 \end{center}
\begin{quote}
Constructs from conformal geometry are important  in low dimensional gravity models, while in higher dimensions the  higher curvature interactions of Lovelock gravity are similarly prominent.
Considering conformal invariance in the context of Lovelock gravity leads to natural, higher-curvature generalizations of the Weyl, Schouten, Cotton and Bach tensors, with properties that straightforwardly extend those of  their familiar counterparts.  As a first application, we introduce a new set of conformally invariant gravity theories in $D=4k$ dimensions, based on the squares of the higher curvature Weyl tensors.
 \vfill
\vskip 2.mm
\end{quote}
\end{titlepage}

\section{Introduction}

In this paper we introduce conformally invariant tensors, and related quantities, associated with the higher curvature interaction terms of Lovelock gravity theories \cite{Lovelock:1971yv}.   This study can be motivated in the following way.  The standard constructs of conformal geometry play key roles in a number of interesting gravity models.  For example, the gravitational Chern-Simons interaction in $D=3$ topologically massive gravity \cite{Deser:1981wh} contributes a term to the field equations that is proportional to the Cotton tensor, whose vanishing is a necessary and sufficient condition for conformal flatness.  A second example in $D=3$ is new massive gravity \cite{Bergshoeff:2009hq}, where the key higher curvature interaction involves the Schouten tensor (see {\it e.g.} the discussion in \cite{Bergshoeff:2010ad}), which characterizes the difference between the Riemann and Weyl tensors.  In $D=4$ renewed attention has been drawn \cite{Maldacena:2011mk} to conformal (Weyl)$^{2}$ gravity, where the equations of motion are proportional to the Bach tensor, which vanishes on all metrics that are conformally related to an Einstein metric. 

Lovelock theories are extensions of general relativity to dimensions greater than four that include a certain limited set of higher curvature interactions.
Lovelock theories are distinguished amongst the much larger class of all 
higher curvature theories by having field equations that depend only on powers of the Riemann tensor, and not on its derivatives.  Importantly, Lovelock theories also have constant curvature vacua with ghost-free excitation spectra \cite{Zwiebach:1985uq}, although not all vacua share this property \cite{Boulware:1985wk}.  Furthermore, many properties of black hole solutions to Lovelock theories, including explicit solutions in the simplest nontrivial cases,  are accessible through analytic means \cite{Boulware:1985wk,Wheeler:1985nh,Wiltshire:1985us,Wheeler:1985qd,Myers:1988ze,Jacobson:1993xs,Cai:2001dz} (see the reviews \cite{Charmousis:2008kc,Garraffo:2008hu,Padmanabhan:2013xyr} for further references).
These favorable properties of  Lovelock  theories have led to their use as testing grounds in studies of a wide variety of physical phenomena, such as brane world models (see \cite{Charmousis:2008kc} and references therein), holographic fluid dynamics (see the review \cite{Cremonini:2011iq} and references therein),
applications of gauge/gravity duality to condensed matter systems \cite{Gregory:2009fj,Pan:2009xa,Pan:2010at,Cai:2010cv,Barclay:2010up,Barclay:2010nm} and entanglement entropy \cite{deBoer:2011wk,Hung:2011xb,Ogawa:2011fw}.  

There is substantial interest to find gravitational models in higher dimensions that realize at least some of the key features of the lower dimensional models mentioned in the first paragraph.  For example, certain properties of $D=3$ new massive gravity have been successfully extended to higher dimensions in the `quasi-topological gravity' models of  \cite{Oliva:2010eb,Myers:2010ru,Oliva:2010zd}, while a  $D=6$ conformal gravity theory was constructed in  \cite{Lu:2013hx} that has all conformally Einstein metrics as solutions, as in $D=4$ (Weyl)$^2$ gravity.   The (Weyl)$^{2}$ interaction term is also a key ingredient in critical gravity models \cite{Lu:2011zk,Lu:2011ks}, which fine-tune higher curvature couplings to make the added gravitational degrees of physically acceptable.

Given the extensive use made of  conformal tensors in low dimensional gravity models, and the similar importance  of Lovelock interactions in higher dimensions, it makes sense to try to combine the two and obtain new tools for gravitational model building.   We will see that such a combination, in fact, arises quite naturally.  The $k$th order Lovelock interaction term has associated with it a conformally invariant tensor first introduced by Kulkarni in the early 1970's \cite{kulkarni}, which we call the Weyl$^{(k)}$ tensor,  that is likewise $k$th order in the curvature.  We will see that Schouten$^{(k)}$ and Cotton$^{(k)}$ tensors can naturally be included in this construction as well.  All have properties that straightforwardly generalize those of their familiar $k=1$ counterparts.
As a first application of these new tools, we explore the properties of  a new conformal (Weyl$^{(k)}$)$^2$ gravity theory in $D=4k$, which has the vanishing of a Bach$^{(k)}$ tensor as its equation of motion and is solved by a generalized class of Einstein$^{(k)}$ metrics.

The paper is organized as follows.  In section \ref{rlsection}, we present the higher curvature interactions of Lovelock gravity theories in terms of Riemann$^{(k)}$ tensors.  These are anti-symmetrized products of Riemann tensors that satisfy symmetries and Bianchi identities that naturally extend those satisfied by the Riemann tensor itself.  Section \ref{weylsection} reviews the definitions, important properties, and inter-relations between certain basic  constructs from conformal geometry, the Weyl, Schouten and Cotton tensors.  Section \ref{square} presents key features of $D=4$ conformal (Weyl)$^2$ gravity, including the statement of the equations of motion in terms of the Bach tensor.  In section \ref{weylksection}, we define (following \cite{kulkarni}) the higher curvature Weyl$^{(k)}$ tensor as the trace free piece of the Riemann$^{(k)}$ tensor and study its properties, along with those of the associated Shouten$^{(k)}$ and Cotton$^{(k)}$ tensors.  In section \ref{newgravity}, we present (Weyl$^{(k)}$)$^2$ gravity which is conformally invariant in $D=4k$, and has solutions that generalize the Einstein metrics of the $k=1$ case.  Finally, section \ref{conclude} provides brief concluding remarks and directions for future investigation.

\section{Lovelock gravity and Riemann$^{(k)}$ tensors}\label{rlsection}

In Lovelock gravity  \cite{Lovelock:1971yv} a particular set of higher curvature interactions terms is added to general relativity.  The action is given by
\begin{equation}\label{action}
S = \int d^Dx\sqrt{-g}\,\sum_{k=0}^{k_D} a_k \calrk
\end{equation}
with the scalar $\calrk$ being $k$th order in the curvature.
In particular, one has $\calr^{(0)}=1$ giving the cosmological term, $\calr^{(1)}=R$ the Einstein term, and $\calr^{(2)}={1\over 6}\left(R_{abcd}R^{abcd}-4R_{ab}R^{ab}+R^2\right)$
the Gauss-Bonnet term.  The scalars $\calrk$ are dimensionally continued Euler densities \cite{Zumino:1985dp}. The integral of $\calrk$ over a closed manifold in $D=2k$ is proportional to its topologically invariant Euler character.  It contributes to the dynamics of Lovelock gravity only when `dimensionally continued' to $D>2k$.  This fixes the upper bound $k_D=[(D-1)/2]$ on the interaction terms contributing to the action.  Finally, the constants $a_k$ appearing in (\ref{action}) are couplings having dimensions 
$(length)^{2k-D}$ respectively.

In the case $k=1$, the scalar curvature can be written as a double trace of the Riemann tensor $R=R_{ab}{}^{ab}$.  A similar expression exists for all the scalars $\calrk$ with $k\ge 1$, in terms of a totally anti-symmetrized product of $k$ Riemann tensors, which we will call the Riemann$^{(k)}$ tensor, given by
\begin{equation}\label{rltensor}
\calrk_{a_1b_1\dots a_kb_k}{}^{c_1d_1\dots c_kd_k} \equiv R_{[a_1b_1}{}^{[c_1d_1}R_{a_2b_2}{}^{c_2d_2}\cdots R_{a_kb_k]}{}^{c_kd_k]}.
\end{equation}
The Riemann$^{(1)}$ tensor is simply the Riemann tensor itself.  
General properties of these tensors were studied in \cite{kulkarni} and their importance in Lovelock gravity was explored in \cite{Kastor:2012se} (see also \cite{Labbi:2007dr,Dadhich:2012cv}).  
The Riemann$^{(k)}$ tensor vanishes identically for $D<2k$, as a consequence of the anti-symmetrizations in (\ref{rltensor}).
For $D\ge 2k$ its symmetries straightforwardly generalize those of the Riemann tensor, 
\begin{equation}\label{symmetries}
\calrk_{a_1\dots a_{2k}b_1\dots b_{2k}} = \calrk_{[a_1\dots a_{2k}]b_1\dots b_{2k}}=\calrk_{a_1\dots a_{2k}[b_1\dots b_{2k}]}=\calrk_{b_1\dots b_{2k}a_1\dots a_{2k}}
\end{equation}
and it likewise satisfies algebraic and differential Bianchi identities analogous to the $k=1$ case, 
\begin{eqnarray}\label{abianchi}
\calrk_{[a_1\dots a_{2k}b_1]} {}^{b_2\dots b_{2k}} &=& 0\\
\nabla_{[c}^{}\calrk_{a_1\dots a_{2k}]}{}^{b_1\dots b_{2k}}&=&0\label{bianchi}.
\end{eqnarray}
One can take the simultaneous trace of a Riemann$^{(k)}$ tensor over $2k-n$ pairs of indices with $0\le n\le2k$ giving the quantities
$\calrk_{a_1\dots a_{n}}{}^{b_1\dots b_{n}}= \calrk_{a_1\dots a_{n}c_1\dots c_{2k-n}}{}^{b_1\dots b_{n}c_1\dots c_{2k-n}}$.
The scalar Lovelock interactions $\calrk$ in (\ref{action}) are obtained by contracting over all $2k$ sets of indices.
The equations of motion obtained by varying the  Lovelock action (\ref{action})  can be written as a sum 
\begin{equation}
 \sum_{k=0}^{k_D} a_k \calg^{(k)a}{}_b=0
 \end{equation}
where the Einstein-like at  $k$th order in the curvature is given by $ \calg^{(k)}_a{}^b=k\calr^{(k)}_a{}^b-(1/2)\delta^a_b\calrk$.
The conservation law $\nabla_a \calg^{(k)a}{}_b=0$ for these tensors  follows from contracting $2k$ pairs of indices in the Bianchi identity (\ref{bianchi}) for the Riemann$^{(k)}$ tensor.

Riemann$^{(k)}$ tensors have interesting properties in  `relatively' low dimensions'  \cite{Kastor:2012se}.  It is clear from (\ref{rltensor}) that in $D=2k$ the Riemann$^{(k)}$ has only a single independent component and is therefore fully determined by the scalar $\calrk$.  It is similarly determined in terms of traces (over fewer indices) for all $D<4k$.   
In particular, in $D=2k+1$ which is the lowest dimension where the interaction $\calrk$ is nontrivial, the Riemann$^{(k)}$ tensor is determined by the Ricci-like tensor $\calr^{(k)}_a{}^b$.  In this case, it is interesting to consider the `pure' $k$th order Lovelock theory such that  only $a_k\neq 0$.  The  equations of motion for this theory are $\calr^{(k)}_a{}^b=0$, which in turn implies that the Riemann$^{(k)}$ tensor vanishes.   For $k=1$, this reduces to the well known behavior of Einstein gravity in $D=3$, where all solutions to the equations of motion are locally flat \cite{Deser:1983tn}.  We can say that all solutions to pure $k$th order Lovelock gravity in $D=2k+1$ are Riemann$^{(k)}$ flat or simply `$k$-flat'.

There turn out to be interesting $k$-flat spacetimes that are not flat  \cite{Kastor:2012se}.  The static spherically symmetric solutions to pure $k$th order Lovelock gravity in $D=2k+1$ are characterized by missing solid angle 
\begin{equation}
ds_{2k+1}^2 = -dt^2 +dr^2 + \alpha^2r^2 d\Omega_{2k-1}^2 ~.
\end{equation}
For $\alpha\neq 1$ the spacetime has non-trivial Riemann curvature, but vanishing Riemann$^{(k)}$ tensor.  This generalizes the $k=1$ case of conical, missing angle spaces as solutions of Einstein gravity in $D=3$ \cite{Deser:1983tn}.
Finally, note that in any dimension we can ask  whether there are spacetimes which are $k$-flat without being flat, or more generally $k$-flat without being $k^\prime$-flat for $k^\prime<k$.  Examples of $k$-flat spacetimes may easily be constructed using product metrics, {\it e.g.} a $D$ dimensional spacetime having one  arbitrarily curved factor with dimension $D_1<2k$ and a second flat factor of dimension $D_2=D-D_1$ will be $k$-flat.  In the case of $1$-flat ({\it i.e.} flat) spacetimes, there is a unique local model for a flat metric, namely the Minkowski metric.  The product example shows that this will not be the case for $k$-flat spacetimes, and that the classification of $k$-flat spacetimes will be non-trivial.

\section{Weyl, Schouten and Cotton tensors}\label{weylsection}

We are interested in finding higher curvature, Lovelock analogues of the conformally invariant, trace free Weyl tensor and certain related geometric constructs, namely the Schouten, Cotton and Bach tensors.   In this section we will review the definitions and basic properties of these quantities, as well as the relations between them.
We start with the  Weyl tensor, which can be written as 
$W_{ab}{}^{cd} = R_{ab}{}^{cd}  - 4\delta_{[a}^{[c}S§_{b]}^{}{}^{d]}$,
with the difference between the Riemann and Weyl tensors given in terms of the Schouten tensor  
\begin{equation}\label{schouten}
S_a{}^b = {1\over D-2}\left (R_a{}^b -  {1\over 2(D-1)}\delta_a^b R\right ) ~.
\end{equation}
The Weyl tensor shares the symmetries of the Riemann tensor, 
$W_{abcd}=W_{[ab]cd}=W_{ab[cd]}=W_{cdab}$,
and also satisfies the algebraic Bianchi identity $W_{[abc]d}=0$.

From the differential Bianchi identity  $\nabla_{[a}R_{bc]}{}^{de}=0$ it follows that the Weyl tensor satisfies
$\nabla_{[a}W_{bc]}{}^{de}=-4 \nabla_{[a}^{}\delta_{b}^{[d}S^{}_{c]}{}^{e]}$.
Further relations are obtained by  taking traces.  Contracting one pair of indices gives a formula for the divergence of the Weyl tensor 
$\nabla_c  W_{ab}{}^{cd} =(D-3) C_{ab}{}^d -2 C_{c[a}{}^c\delta_{b]}^d$
where the Cotton tensor is defined as the curl of the Schouten tensor
\begin{equation}\label{cotton}
C_{ab}{}^c = 2\nabla_{[a}S_{b]}{}^c ~.
\end{equation}
Further contracting a second pair of indices shows that the Cotton tensor is traceless $C_{ab}{}^b=0$.  Taking this into account,  the divergence of the Weyl tensor then reduces to 
\begin{equation}\label{weyldiv}
\nabla_c  W_{ab}{}^{cd} =(D-3) C_{ab}{}^d ~.
\end{equation}
The Cotton tensor itself can be shown to be divergenceless $\nabla_c\, C_{ab}{}^c=0$.
As noted in the introduction, the Schouten and Cotton tensors play important roles in three dimensional massive gravity theories \cite{Deser:1981wh,Bergshoeff:2009hq}.

The Weyl tensor is defined only for $D\ge 3$, as one sees from the expression for the Schouten tensor.   Moreover, in $D=3$ the Weyl tensor vanishes identically. This can be seen from considering the relation\footnote{The  multi-index delta symbol, which we will make extensive use of below,  denotes an antisymmetrized product of Kronecker delta functions with overall unit strength, so that $\delta_{a_1\dots a_n}^{b_1\dots b_n} = \delta_{[a_1}^{b_1}\dots \delta_{a_n]}^{b_n}=\delta_{a_1}^{[b_1}\dots \delta_{a_n}^{b_n]}$.}
$\delta_{abgh}^{cdef}\, W_{ef}{}^{gh}= (1/ 6)W_{ab}{}^{cd}$
which is valid in any dimension, following from the tracelessness of the Weyl tensor.   In $D=3$ the left hand side necessarily vanishes due to the anti-symmetrization over four coordinate indices, giving the result\footnote{This is an example of  the method given by Lovelock  in \cite{lovelock} for proving what he called `dimension dependent identities'.}.   The fact that the curvature  tensor has no trace free piece in $D\le 3$ is due respectively to the absence of curvature in $D=1$, and that the Riemann tensor is determined by its traces in $D=1$ and $D=2$.  

An efficient way to demonstrate the conformal invariance of the Weyl tensor is presented in \cite{kulkarni}.
Consider  a tensor $A_{ab}{}^{cd}$ with the algebraic symmetries of the Riemann tensor $A_{abcd}=A_{[ab]cd}=A_{ab[cd]}=A_{cdab}$ and denote its traces by $A_a{}^c=A_{ab}{}^{cb}$ and $A=A_a{}^a$.  Its tracefree part is given by
\begin{equation}\label{gencon}
 A^{(t)}_{ab}{}^{cd} = A_{ab}{}^{cd}-{4\over D-2} \delta_{[a}^{[c} A^{}_{b]}{}^{d]} + {2\over (D-1)(D-2)}\delta_{ab}^{cd} A , 
\end{equation}
which reproduces the Weyl tensor if $A_{ab}{}^{cd}=R_{ab}{}^{cd}$.
If we take a tensor with the pure trace form $A_{ab}{}^{cd}= \delta_{[a}^{[c} \Lambda^{}_{b]}{}^{d]}$ where $\Lambda_{ab}=\Lambda_{ba}$, then  this satisfies the algebraic symmetries of the Riemann tensor, and one finds that its traceless part  $A^{(t)}_{ab}{}^{cd} $ vanishes identically.
Now consider also the Riemann tensor of a conformally related metric $\tilde g_{ab}= e^{2f} g_{ab}$.  This is related to the Riemann tensor of the original metric according to
\begin{equation}\label{conformal}
\tilde R_{ab}{}^{cd} = e^{-2f} \left(R_{ab}{}^{cd} + \delta_{[a}^{[c} \Lambda^{}_{b]}{}^{d]}\right)
\end{equation}
where $\Lambda_{a}{}^b=4\nabla_a\nabla^b f+4(\nabla_a f)\nabla^b f -2\delta_a^b(\nabla_cf)\nabla^c f$
and one sees that $\Lambda_{ab}$ is symmetric.
By the reasoning presented above, the second term on the right hand side of (\ref{conformal}) 
will not contribute to the Weyl tensor for the transformed metric, which is therefore given by
\begin{equation}
\tilde W_{ab}{}^{cd}=e^{-2f} W_{ab}{}^{cd}~.
\end{equation}
The conformal transformation of the Cotton tensor, derived in appendix (\ref{allcotton}), is given by
\begin{equation}\label{cottontrans}
\tilde C_{ab}{}^c=e^{-2f}\left( C_{ab}{}^c-W_{ab}{}^{cd}\nabla_df\right)~.
\end{equation}
One sees that the Cotton tensor is conformally invariant if the Weyl tensor vanishes.  This property underlies the special significance of the Cotton tensor in $D=3$ where the Weyl tensor vanishes identically.

The Weyl tensor vanishes for flat spacetime and therefore also in spacetimes that are conformally flat.  It follows that $W_{ab}{}^{cd}=0$ is a nessecary condition for a spacetime to be conformally flat.  
In dimensions $D\ge 4$, it can be shown that this condition is also a sufficient.  Lower dimensions are special cases.  As we have seen, the Weyl tensor vanishes identically in $D=3$.  However, it then follows from (\ref{cottontrans}) that the Cotton tensor is conformally invariant, and it can be shown that $C_{ab}{}^c=0$ is both a necessary and sufficient condition for conformal flatness in $D=3$.  It is well known that all metrics are locally conformally flat in $D=2$, while all metrics are flat in $D=1$.

\section{Conformal (Weyl)$^2$ gravity in $D=4$}\label{square}

There is a unique conformally invariant gravity theory in $D=4$ with action given by the square of the Weyl tensor
\begin{equation}
S=\int d^4x\sqrt{-g}\, W_{ab}{}^{cd}W_{cd}{}^{ab}.
\end{equation}
The equations of motion for this theory may be written as $B_a{}^b=0$ where
\begin{equation}\label{bach}
B_a{}^b= (\nabla^d\nabla_c +{1\over 2}R_c{}^d)W_{ad}{}^{bc} ~.
\end{equation}
is the Bach tensor, which is symmetric, traceless and transforms as $\tilde B_a{}^b=e^{-4f}B_a{}^b$ under a conformal transformation $\tilde g_{ab}=e^{2f} g_{ab}$.
It follows that if a given metric satisfies the equations of motion, then so do conformally related metrics.
Writing the Bach tensor in the form given in (\ref{bach}) requires a quadratic identity for the Weyl tensor which holds only in $D=4$
\begin{equation}\label{quadweyl}
W_{cd}{}^{be}W_{ae}^{cd}={1\over 4}\delta_a^b W_{cd}{}^{ef}W_{ef}{}^{cd}
\end{equation}
This is another example of a dimension dependent identity \cite{lovelock}.  It is proved by considering the quantity 
$\delta_{aghij}^{bcdef}W_{cd}{}^{gh}W_{ef}{}^{ij}= (1/ 5)\left(\delta_a^b W_{cd}{}^{ef}W_{ef}{}^{cd}- 4W_{ae}{}^{cd}W_{cd}{}^{be}\right)$
which vanishes identically in $D=4$ due to the antisymmetrization over 5 coordinate indices on the left hand side.  

It is straightforward to check that the equations of motion $B_a{}^b=0$ are solved by any metric having $R_{ab}=\alpha g_{ab}$ with $\alpha$ constant.  These are known as Einstein metrics.  First one sees that with this form for the Ricci tensor the contraction $R_c{}^dW_{ad}{}^{bc}$ in (\ref{bach})  is proportional to the trace of the Weyl tensor, which vanishes by definition.  The remaining term in (\ref{bach}) involves the divergence of the Weyl tensor and can be rewritten in terms of the Cotton tensor via equation (\ref{weyldiv}).   However, the Cotton tensor (\ref{cotton}) vanishes for Einstein metrics, completing the proof.
It has been shown recently \cite{Maldacena:2011mk} for asymptotically deSitter spacetimes (or Euclidean asymptotically AdS spaces) that the set of  Einstein metrics may be selected out of the full set of solutions to conformal gravity by means of boundary conditions in the asymptotic region\footnote{See also the earlier related results in \cite{Miskovic:2009bm}.}.  Non-conformally-Einstein solutions to conformal gravity in $D=4$ have also recently been studied \cite{Liu:2013fna}.

\section{Lovelock counterparts of the Weyl, Schouten and Cotton tensors}\label{weylksection}

We now consider conformal tensors related to the higher curvature Lovelock interactions.
The symmetry and other properties of  the Riemann tensor, which entered the discussion of conformal tensors in the last section, all extend to the general case of 
Riemann${}^{(k)}$ tensors with $k\ge 1$.
It is therefore natural to continue the Riemann$^{(k)}$ construction one step further and build associated  trace free tensors, which we call Weyl${}^{(k)}$ tensors.   We will see that the Weyl$^{(k)}$ tensors with $k\ge 1$ are conformally invariant and that one can further define at each higher curvature order Schouten$^{(k)}$, Cotton$^{(k)}$ and Bach$^{(k)}$ tensors that have properties and inter-relations analogous to their $k=1$ counterparts.

The Weyl$^{(k)}$ tensors were first constructed by Kulkarni in  reference \cite{kulkarni}.  Kulkarni
began by considering tensors $A_{a_1\dots a_n}{}^{b_1\dots b_n}$ that have the algebraic symmetries in (\ref{symmetries}) characteristic of the Riemann${}^{(k)}$ tensors.
The associated trace free tensor then has the form 
\begin{equation}\label{bigweyl}
A^{(t)}_{a_1\dots a_n}{}^{b_1\dots b_n}=A_{a_1\dots a_n}{}^{b_1\dots b_n} + \sum_{p=1}^n\, \alpha_p\, \delta_{[a_1\dots a_p}^{[b_1\dots b_p}
A^{}_{a_{p+1}\dots a_n]}{}^{b_{p+1}\dots b_n]}
\end{equation}
where $A_{a_1\dots a_m}{}^{b_1\dots b_m}\equiv A_{a_1\dots a_mc_1\dots c_{n-m}}{}^{b_1\dots b_mc_1\dots c_{n-m}}$
with $0\le m < n$ are (multiple) traces of the original tensor.  By requiring that the trace of $A^{(t)}_{a_1\dots a_n}{}^{b_1\dots b_n}$ should vanish, one finds that the coefficients in (\ref{bigweyl}) are given recursively by the formula
\begin{equation}\label{bigweyl2}
\alpha_{p+1} = -{(n-p)^2\over (p+1)(D+p-2(n-1))}\alpha_p
\end{equation}
with $\alpha_0\equiv 1$.  
In order to establish the conformal invariance of the Weyl$^{(k)}$ tensor, defined below, it is important to note that if  a tensor $A_{a_1\dots a_n}{}^{b_1\dots b_n}$ has the pure trace form
\begin{equation}
A_{a_1\dots a_n}{}^{b_1\dots b_n}= \delta_{[a_1}^{[b_1} \Lambda^{}_{a_2\dots a_n]}{}^{b_2\dots b_n]} 
\end{equation}
then its trace free part $A^{(t)}_{a_1\dots a_n}{}^{b_1\dots b_n}$ necessarily vanishes.  
The Weyl${}^{(k)}$ tensor \cite{kulkarni}  is now defined as the traceless part of the Riemann${}^{(k)}$ tensor, so that one has
\begin{eqnarray}
\calw^{(k)}_{a_1\dots a_{2k}}{}^{b_1\dots b_{2k}}&=&\calr^{(k,t)}_{a_1\dots a_{2k}}{}^{b_1\dots b_{2k}}\\
&=&	\calr^{(k)}_{a_1\dots a_{2k}}{}^{b_1\dots b_{2k}} + \sum_{p=1}^{2k}\, \alpha_p\, \delta_{[a_1\dots a_p}^{[b_1\dots b_p}
\calrk_{a_{p+1}\dots a_{2k}]}{}^{b_{p+1}\dots b_{2k}]} ~.
\end{eqnarray}
To understand the properties of the Weyl$^{(k)}$ tensors, it will be helpful to evaluate the recursive formula (\ref{bigweyl2}) for the coefficients $\alpha_p$ to obtain the explicit expressions
\begin{equation}\label{coefficients}
\alpha_p = \left({(2k)!\over (2k-p)!} \right)^2\, {(-1)^p(D-(4k-1))!\over p!(D-(4k-p-1))!}~.
\end{equation}

For $k=1$ these formulas correctly reproduce the ordinary Weyl tensor. 
Some basic properties of the Weyl$^{(k)}$ tensors are as follows.
Singular factors in the coefficients 
$\alpha_p$ indicate that  that the Weyl$^{(k)}$ tensor is defined only for $D\ge 4k-1$.
Moreover, one can show that the Weyl${}^{(k)}$ tensor vanishes identically in $D=4k-1$ by
considering the relation
\begin{equation}
\delta_{a_1\dots a_{2k}d_1\dots d_{2k}}^{b_1\dots b_{2k}c_1\dots c_{2k}}\, \calw^{(k)}_{c_1\dots c_{2k}}{}^{d_1\dots d_{2k}}
={(2k!)^2\over 4k!}\, \calw^{(k)}_{a_1\dots a_{2k}}{}^{b_1\dots b_{2k}}
\end{equation}
which holds in any dimension.  The left hand side, however, vanishes identically in $D=4k-1$ due to the antisymmetrization over a total of $4k$ coordinate indices, leading to the result.  Hence the Weyl$^{(k)}$ tensor is non-trivial only for $D\ge 4k$.  This result also follows from the observations in \cite{Kastor:2012se} that the Riemann$^{(k)}$ vanishes identically for $D<2k$, is fully determined by its traces for $2k\le D <4k$, and hence has a non-trivial trace free part only in dimensions $D\ge 4k$.

The conformal properties of  Weyl${}^{(k)}$ tensors were also established in \cite{kulkarni}.  It follows from the definition of the Riemann${}^{(k)}$ tensor (\ref{rltensor})  together with the conformal transformation of the Riemann tensor in (\ref{conformal}), that the action of a conformal transformation on the Riemann${}^{(k)}$ is given by
\begin{equation}\label{kconf}
\tilde\calr{}^{(k)}_{a_1\dots a_{2k}}{}^{b_1\dots b_{2k}}=e^{-2kf}\left(\calrk_{a_1\dots a_{2k}}{}^{b_1\dots b_{2k}} + \delta_{[a_1}^{[b_1}\Lambda^{}_{a_2\dots a_{2k}]}{}^{b_2\dots b_{2k}]}\right)
\end{equation}
where $\Lambda^{}_{a_1\dots a_{2k-1}}{}^{b_1\dots b_{2k-1}}$ is a tensor having the symmetries 
\begin{equation}
\Lambda_{a_1\dots a_{2k-1}b_1\dots b_{2k-1}}=\Lambda_{[a_1\dots a_{2k-1}]b_1\dots b_{2k-1}}=\Lambda_{a_1\dots a_{2k-1}[b_1\dots b_{2k-1}]}=\Lambda_{b_1\dots b_{2k-1}a_1\dots a_{2k-1}}
\end{equation}
whose precise form will not be needed.
It then follows from an argument given above
that the second term on the right hand side of (\ref{kconf}) does not contribute to the 
trace free, transformed Weyl$^{(k)}$ tensor, and that therefore one has
\begin{equation}
\tilde\calw^{(k)}_{a_1\dots a_{2k}}{}^{b_1\dots b_{2k}}=e^{-2kf}\, \calw^{(k)}_{a_1\dots a_{2k}}{}^{b_1\dots b_{2k}}.
\end{equation}
We see then that the Weyl$^{(k)}$ tensors constitute an interesting class of conformally invariant tensors in higher dimensions that are naturally associated with the geometric constructs of Lovelock gravity theories.

We now continue the investigation of higher order conformal tensors beyond the point where reference \cite{kulkarni} leaves off, introducing Schouten$^{(k)}$ and Cotton$^{(k)}$ tensors that also  generalize the $k=1$ case.  It will follow from the Bianchi identity (\ref{bianchi}) for the  Riemann$^{(k)}$ tensor,
that the  Weyl$^{(k)}$, Schouten$^{(k)}$ and Cotton$^{(k)}$  tensors with $k>1$ are inter-related in the same manner as their ordinary $k=1$ counterparts.
We may begin by writing the Weyl$^{(k)}$ tensor as
\begin{equation}
\calw^{(k)}_{a_1\dots a_{2k}}{}^{b_1\dots b_{2k}}=\calr^{(k)}_{a_1\dots a_{2k}}{}^{b_1\dots b_{2k}}
-(2k)^2\delta_{[a_1}^{[b_1}\,  \cals^{(k)}_{a_2\dots a_{2k}]}{}^{b_2\dots b_{2k}]}
\end{equation}
in terms of  the Schouten$^{(k)}$ tensor 
\begin{equation}
\cals^{(k)} _{a_1\dots a_{2k-1}}{}^{b_1\dots b_{2k-1}} = -{1\over (2k)^2}\sum_{p=0}^{2k-1}\alpha_{p+1}
\delta_{[a_1\dots a_p}^{[b_1\dots b_p}\calrk_{a_{p+1}\dots a_{2k-1}]}{}^{b_{p+1}\dots b_{2k-1}]}
\end{equation}
where the coefficients are those given in (\ref{coefficients}).  The Schouten$^{(k)}$ tensors have the symmetries
\begin{equation}
\cals^{(k)}_{a_1\dots a_{2k-1}b_1\dots b_{2k-1}}=\cals^{(k)}_{[a_1\dots a_{2k-1}]b_1\dots b_{2k-1}}=\cals^{(k)}_{a_1\dots a_{2k-1}[b_1\dots b_{2k-1}]}=\cals^{(k)}_{b_1\dots b_{2k-1}a_1\dots a_{2k-1}}
\end{equation}
The Cotton$^{(k)}$ tensor can now be defined in analogy with equation (\ref{cotton}) as the curl of the Schouten$^{(k)}$ tensor
\begin{equation}
\calc^{(k)}_{a_1\dots a_{2k}}{}^{b_1\dots b_{2k-1}} = 2k\, \nabla^{}_{[a_1}\cals^{(k)}_{a_2\dots a_{2k}]}{}^{b_1\dots b_{2k-1}}~.
\end{equation}

Due to the Bianchi identity (\ref{bianchi}) for the Riemann$^{(k)}$ tensor the curl of the Weyl$^{(k)}$ tensor is determined by the Cotton$^{(k)}$ tensor
$\nabla_{[a_1}^{}\calw^{(k)}_{a_2\dots a_{2k+1}]}{}^{b_1\dots b_{2k}}=-2k\, \calc^{(k)}_{[a_1\dots a_{2k}}{}^{[b_1\dots b_{2k-1}}\delta_{a_{2k+1}]}^{b_{2k}]}$.
Contracting one set of upper and lower indices then leads to a relation for the divergence of the Weyl$^{(k)}$ tensor
\begin{eqnarray}\label{divergence}
\nabla_c \calw^{(k)}_{a_1\dots a_{2k}}{}^{cb_1\dots b_{2k-1}} =&& (D-(4k-1))\calc^{(k)}_{a_1\dots a_{2k}}{}^{b_1\dots b_{2k-1}}\\
&& \nonumber -(2k)(2k-1)\calc^{(k)}_{c[a_1\dots a_{2k-1}}{}^{c[b_1\dots b_{2k-2}}\delta_{a_{2k}]}^{b_{2k-1}]}
\end{eqnarray}
The trace of the Cotton$^{(k)}$ tensor, which comes into the second term on the right hand side, can ultimately be seen to vanish by taking further traces of equation (\ref{divergence}) giving $\calc^{(k)}_{a_1\dots a_{2k-1}c}{}^{b_1\dots b_{2k-2}c}=0$.
This finally yields a result for the divergence of the Weyl$^{(k)}$ tensor that generalizes  equation (\ref{weyldiv}) in the $k=1$ case
\begin{equation}
\nabla_c \calw^{(k)}_{a_1\dots a_{2k}}{}^{cb_1\dots b_{2k-1}} = \left(D-(4k-1)\right)\, \calc^{(k)}_{a_1\dots a_{2k}}{}^{b_1\dots b_{2k-1}}
\end{equation}
The factor of $D-(4k-1)$ on the right hand side ensures the consistency of the Cotton$^{(k)}$ tensor being non-vanishing in $D=4k-1$, while the Weyl$^{(k)}$ tensor vanishes identically, but the Cotton tensor can be non-vanishing.
Finally, the conformal transformation of the Cotton$^{(k)}$ tensor is found in appendix (\ref{allcotton}) to be given by
\begin{equation}\label{cottonkconformal}
\tilde\calc^{(k)}_{a_1\dots a_{2k}}{}^{b_1\dots b_{2k-1}} = e^{-2kf}\left ( \calc^{(k)}_{a_1\dots a_{2k}}{}^{b_1\dots b_{2k-1}} -
\calw^{(k)}_{a_1\dots a_{2k}}{}^{b_1\dots b_{2k-1}c}\nabla_c f\right )
\end{equation}
which generalizes equation (\ref{cottontrans}).
We see from this that the Cotton$^{(k)}$ tensor is conformally invariant if the Weyl$^{(k)}$ tensor vanishes.  In particular, this will be the case in $D=4k-1$ where the Weyl$^{(k)}$ tensor vanishes identically.

The property of conformal flatness is of considerable interest in both physical and mathematical contexts.  As noted above $W_{ab}{}^{cd}=0$ is a necessary and sufficient condition for conformal flatness in dimensions $D\ge 4$, while the condition $C_{ab}{}^c=0$ plays a similar role in $D=3$.
We have defined the property of $k$-flatness above as the vanishing of the Riemann$^{(k)}$ tensor and discussed some examples of spacetimes that are $k$-flat for  $k>1$, without being flat.  With corresponding `conformal$^{(k)}$' tensors now in hand, we can speculate about conditions for `conformal k-flatness', whether a given spacetime may be related to a k-flat one via a conformal transformation.  It is obvious that vanishing of the Weyl$^{(k)}$ tensor is a necessary condition for $k$-flatness in dimension $D\ge 4$.  It is natural to conjecture that this is also a sufficient condition.  A similar conjecture regarding conformal $k$-flatness in $D=4k-1$ and the vanishing of the Cotton$^{(k)}$ tensor seems natural as well.  

The analogue of $D=2$, where all geometries are locally conformally flat, appears to be $D=2k$.  In this case the Riemann$^{(k)}$ tensor has a single nontrivial component, and one can envision this as being set to zero by the single degree of freedom in a conformal transformation.  
The range of dimensions $2k<D<4k-1$, however, is empty in the $k=1$ case, and there is no obvious conjecture to make regarding necessary and sufficient conditions for conformal $k$-flatness for dimensions in this range.  The Riemann$^{(k)}$ tensor is determined by its traces in these dimensions, but none of the conformal tensors introduced above are defined.  It is intriguing to consider generalizing the proofs regarding conformal flatness for $D=2$, $D=3$ and $D\ge 4$ to the conformally $k$-flat case with $k>1$,  but we will not pursue this here.

\section{Conformal (Weyl$^{(k)}$)$^2$ gravity in $D=4k$}\label{newgravity}

There is no longer  a unique conformally invariant gravitational action in higher dimensions \cite{Deser:1993yx}. 
A unique theory was singled out in $D=6$, from the full three parameter family of conformal gravities, by additionally requiring that it admit all conformally Einstein spacetimes as solutions  \cite{Lu:2013hx}.  Such a procedure could be carried out in still higher dimension as well, although there is no guarantee that this would produce a unique result.  Here, we will present a different set of higher dimensional conformal gravity theories, distinguished by their close association with Lovelock theories.

In section \ref{weylsection} we discussed (Weyl)$^2$ gravity, which is the unique conformally invariant gravity theory in $D=4$.  
Similar conformal gravity theories may be constructed in $D=4k$ dimensions using the square of the Weyl$^{(k)}$ tensors, with actions given by
\begin{equation}\label{kgravity}
S=\int d^{4k}x\sqrt{-g}\,\, \calw^{(k)}_{a_1\dots a_{2k}}{}^{b_1\dots b_{2k}}\, \calw^{(k)}_{b_1\dots b_{2k}}{}^{a_1\dots a_{2k}}.
\end{equation}
Under a conformal transformation $\tilde g_{ab}=e^{2f} g_{ab}$, the two factors of the Weyl$^{(k)}$ tensor in (\ref{kgravity}) will together pick up a factor of $e^{-4kf}$, while the volume element  in $D=4k$ transforms by a compensating factor of $e^{4kf}$, yielding a  conformally invariant action.
One finds that the equation of motion for this theory can be written in terms of the Bach$^{(k)}$ tensor,
\begin{equation}\label{bachk}
\calb^{(k)}_a{}^b= \left ( \calr^{(k-1)}_{c_1\dots c_{2k-2}}{}^{d_1\dots d_{2k-2}}\nabla^{d_{2k-1}}\nabla_{c_{2k-1}}
+{k\over 2} \calr^{(k)}_{c_1\dots c_{2k-1}}{}^{d_1\dots d_{2k-1}}\right)
\calw^{(k)}_{ad_1\dots d_{2k-1}}{}^{bc_1\dots c_{2k-1}} 
\end{equation}
which reduces for $k=1$ to the ordinary Bach tensor (\ref{bach}).  Note that the expression for $B^{(k)}_{ab}$ which one obtains by lowering an index will not be manifestly symmetric.  Strictly speaking, we have shown that the equation of motion is that the symmetric part of the Bach$^{(k)}$ tensor should vanish.  In the $k=1$ case, it can be shown that the Bach tensor is, in fact, symmetric (see {\it e.g.} \cite{bergman}) and that the equation of motion for $D=4$ (Weyl)$^2$ gravity is simply $B_{ab}=0$.  For $k>1$, we similarly expect,
but have not yet been able to show, that the Bach$^{(k)}$ tensor is symmetric and that the equations of motion of $D=4k$ (Weyl$^{(k)}$)$^2$ gravity are simply $\calb^{(k)}_{ab}=0$.
In the following, we will assume that this is the case.  

Writing the equation of motion in this form requires using a quadratic identity for the Weyl$^{(k)}$ tensor that holds in $D=4k$, which is obtained by considering the
\begin{equation}
\delta_{f b_1\dots b_{2k} d_1\dots d_{2k}}^{e a_1\dots a_{2k} c_1\dots c_{2k}}\,
\calw^{(k)}_{a_1\dots a_{2k}}{}^{b_1\dots b_{2k}}\, \calw^{(k)}_{c_1\dots c_{2k}}{}^{d_1\dots d_{2k}}
\end{equation}
On one hand, this quantity vanishes identically in $D=4k$.  On the other hand we can evaluate is explicitly in all dimensions, and thereby arrive at the identity
\begin{equation}
\calw^{(k)}_{a_1\dots a_{2k}}{}^{b_1\dots b_{2k}}\, \calw^{(k)}_{a_1\dots a_{2k}}{}^{b_1\dots b_{2k}}\delta^c_d =
4k\, \calw^{(k)}_{da_2\dots a_{2k}}{}^{b_1\dots b_{2k}}\, \calw^{(k)}_{ca_2\dots a_{2k}}{}^{b_1\dots b_{2k}}\
\end{equation}
generalizing the quadratic identity (\ref{quadweyl}) for the Weyl tensor in $D=4$.

We have seen that all Einstein metrics solve the equations of motion of (Weyl)$^2$ gravity in $D=4$.   The generalization of this statement to all values of $k\ge 1$ is straightforward.
The Ricci tensor for an Einstein metric has the constant form $R_a{}^b=\alpha\, \delta_a^b$.  To obtain solutions to $\calb^{(k)}_a{}^b=0$, the correct generalization is to take the constant form for the first trace of the Riemann$^{(k)}$ tensor
\begin{equation}\label{constant}
 \calr^{(k)}_{a_1\dots a_{2k-1}}{}^{b_1\dots b_{2k-1}}=\alpha\, \delta_{a_1\dots a_{2k-1}}^{b_1\dots b_{2k-1}}~.
 \end{equation}
The term in (\ref{bachk}) involving the contraction of this tensor with the Weyl$^{(k)}$ tensor will then be proportional to a multiple trace of the Weyl$^{(k)}$, which vanishes by definition.  The remaining term in (\ref{bachk}) can be rewritten in terms of the Cotton$^{(k)}$ tensor, which is seen to vanish by noting that (\ref{constant}) implies that the Schouten$^{(k)}$ tensor has a similar constant form.  With the perspective of (Weyl$^{(k)}$)$^2$ in $D=4k$ in mind, it seems reasonable to call spacetimes satisfying the condition (\ref{constant}) Einstein$^{(k)}$ metrics.  They are generalizations of ordinary Einstein spacetimes in the sense that the first trace of the relevant $k$th order curvature tensor has a constant form.

\section{Conclusions} \label{conclude}

We have seen that the geometric structures underlying the higher curvature interactions of Lovelock gravity also provide higher curvature analogues of the tensors of conformal geometry; the Weyl, Schouten, Cotton and Bach tensors.  We have explored the properties and inter-relations of the higher curvature Weyl$^{(k)}$, Schouten$^{(k)}$, Cotton$^{(k)}$ and Bach$^{(k)}$ tensors, and seen that, despite the presence of a cumbersome number of indices, these are straightforward generalizations of familiar results.  
Our hope is that this set of conformal$^{(k)}$ tensors will provide new tools for gravitational model building.  
We have given a first such application by presenting the $D=4k$ conformal (Weyl$^{(k)}$)$^2$ gravity models in section \ref{newgravity}.
From a physics perspective, some next steps would be to investigate what roles conformal$^{(k)}$ tensors can play in contexts such as quasi-topological gravity \cite{Oliva:2010eb,Myers:2010ru,Oliva:2010zd} and critical gravity models \cite{Lu:2011zk,Lu:2011ks}.  From a mathematical perspective, there are potentially interesting questions raised in the text relating to $k$-flatness and conformal $k$-flatness to follow up on.

\subsection*{Acknowledgements}
The author thanks Rob Kusner, Franz Pedit, Sourya Ray and Jennie Traschen for helpful conversations.

\appendix

\section*{Appendix}

\section{Conformal transformation of Cotton tensor}\label{allcotton}

The transformation law (\ref{cottontrans}) for the Cotton tensor under a conformal rescaling of the metric can be obtained as follows.  The action of the covariant derivative for the conformally rescaled metric on a vector field $v^a$ is related to the original covariant derivative according to $\tilde\nabla_av^c = \nabla_av^c +\lambda_{ab}{}^c\, v^b$
where $\lambda_{ab}{}^c= (\nabla_a f) \delta_b^c +(\nabla_b f)\delta_a^c - g_{ab}(\nabla^c f)$.
Note that equation (\ref{weyldiv}) can be equivalently rewritten as 
\begin{equation}
C_{ab}{}^d=-{3\over D-3}\nabla_{[a}W_{bc]}{}^{dc}.
\end{equation}
Writing the transformed Cotton tensor in this way in terms of the transformed Weyl tensor and covariant derivative operator, it then follows that
\begin{equation}\label{confcon}
\tilde C_{ab}{}^d= -{3\over D-3}\left\{ \nabla_{[a}\tilde W_{bc]}{}^{dc} +2\lambda_{e[a}{}^{[c}\, \tilde W_{bc]}{}^{d]e}\right\}  ~.
\end{equation}
It also follows straightforwardly from the transformation law for the Weyl tensor that
\begin{equation}
\nabla_{[a}\tilde W_{bc]}{}^{dc}= -{1\over 3}e^{-2f}\left( (D-3)C_{ab}{}^d +2 W_{ab}{}^{de}\nabla_ef\right)
\end{equation}
while making use of the algebraic Bianchi identity $W_{[abc]}{}^d=0$ for the Weyl tensor and plugging in the expression for $\lambda_{ab}{}^c$  yields
\begin{equation}
2\lambda_{e[a} {}^{[c}\, \tilde W_{bc]}{}^{d]e} = {D-1\over 3} e^{-2f}W_{ab}{}^{de}\nabla_e f ~.
\end{equation}
These combine to give the result
\begin{equation}
\tilde C_{ab}{}^c=e^{-2f}\left( C_{ab}{}^c-W_{ab}{}^{cd}\nabla_df\right)~.
\end{equation}
One sees that if the Weyl tensor vanishes, as it does identically in $D=3$ or in conformally flat spacetimes in $D\ge 4$, then the Cotton tensor is conformally invariant.
The computation in the general case proceeds in a similar manner, making use of the algebraic Bianchi identity for the Weyl$^{(k)}$ tensor and yielding equation (\ref{cottonkconformal}).

\end{document}